\documentstyle[12pt,psfig]{article}
\begin{document}
\renewcommand{\figurename}{\small {\sc Figure}}
\renewcommand{\tablename}{\small{\sc Table}}
\begin{center}
%{\large \bf Searching for $\eta$d and $\eta \alpha$ states from time delay\\}
{\large \bf Small $\eta N$ scattering lengths favour $\eta d$ and 
$\eta \alpha$ states}
\vskip0.7cm
{N. G.~Kelkar$^1$, K. P. Khemchandani$^{2 \dagger}$ and B. K. Jain$^2$\\}
\vskip0.3cm
{$^1$Departamento de Fisica, Universidad de los Andes, Colombia\\
$^2$Department of Physics, University of Mumbai,
Mumbai, India\\
$^{\dagger}$ Departamento de Fisica Teorica and IFIC, 
Centro Mixto Universidad de Valencia-CSIC
Institutos de Investigacion de Paterna, Aptd. 22085, 46071 Valencia, Spain\\
}
\end{center}
\begin{abstract}
Unstable states of the eta meson and the $^3$He nucleus predicted using 
the time delay method were found to be in agreement with a recent claim 
of $\eta$-mesic $^3$He states made by the TAPS collaboration. 
Here, we extend this method to a 
speculative study of the unstable states occurring in the 
$\eta$d and $\eta^4$He elastic scattering.  
The $T$-matrix for $\eta\,^4$He scattering is evaluated within 
the Finite Rank Approximation (FRA) of few body equations. 
For the evaluation of time delay in the $\eta$d case, we use a 
parameterization of an 
existing Faddeev calculation and compare the results with those obtained 
from FRA. With an $\eta N$ scattering length, $a_{\eta N} = (0.42, 0.34)$ fm,  
we find an $\eta d$ unstable bound state around $-16$ MeV, within the 
Faddeev calculation. A similar state within the FRA is found for a low
value of $a_{\eta N}$, namely, $a_{\eta N} = (0.28, 0.19)$ fm. 
The existence of an $\eta ^4$He unstable bound state close to threshold
is hinted by $a_{\eta N} = (0.28, 0.19)$ fm, but is ruled out by large
scattering lengths.

\end{abstract}                                                                
\newpage
\section{Introduction}
More than a decade after the prediction of the $\eta$-mesic nuclei 
\cite{hailiu}, their existence remains a matter of interesting debate 
among nuclear physicists. The root of the debate lies 
in the uncertainty in the knowledge of the elementary eta-nucleon 
($\eta N$) interaction. Though once again, the first prediction of an
attractive $\eta N$ interaction \cite{bhale} (arising basically due 
to the proximity of the $\eta N$ threshold to the $N^* (1535)$ resonance) 
was made back in 1985, an agreement on the magnitude of the attraction 
has not been reached. 
Though most of the
models are constrained by the same sets of data on $\pi N$ elastic scattering
and the $\pi N \rightarrow \eta N$ cross sections, the $\eta N$ scattering
length predicted by different models is very different
due to the unavailability of direct 
experimental information on $\eta N$ elastic scattering. Consequently, the
predictions of possible resonances or unstable bound states of $\eta$ mesons
and nuclei within these models also vary a lot.
In the present article we refer to $\eta$-nucleus states with a negative
binding energy but a finite lifetime (width) as ``unstable bound states".
These states are sometimes also called ``quasibound" states in literature.

Even if the present knowledge of the $\eta N$ interaction is somewhat poor, 
the growing experimental efforts in the past few years could improve the 
understanding in a not so distant future. 
The photoproduction of $\eta$-mesic $^3$He investigated by the TAPS 
collaboration \cite{pfieff}, has indeed proved to be a step forward in this
direction. The total inclusive cross section for the $\gamma \,^3$He 
$\rightarrow \eta X$ reaction was measured at the Mainz Microtron accelerator 
facility using the TAPS calorimeter and an unstable bound state with a 
binding energy of $-4.4 \pm 4.2$ MeV was reported. 
Certain evidence for the existence of the $\eta$-mesic nucleus $^{11}_{\eta}C$, 
was reported in \cite{sokol} and a claim for the light $\eta^4$He 
quasi-bound state was made in \cite{willis} from the study of the cross 
section and tensor analysing power in the $\vec{d} d \rightarrow \, 
^4$He$\,\eta$ reaction. Indirect evidence of the strong $\eta$-nucleus 
attraction was also obtained from data on $\eta$ production in the $p d 
\rightarrow \,^3$He$\,\eta$ \cite{mayer} and $n p \rightarrow d \eta$ 
\cite{npdata} reactions which display large enhancements in the cross 
sections near threshold. More information is expected to be available 
from the ongoing program of the COSY-GEM collaboration \cite{cosygem}.

Since the $\eta^3$He states which we recently predicted 
\cite{wejphysg} from time delay were found to be in agreement with
the TAPS data, we thought it worthwhile to extend the calculations
and search for eta-mesic states in two other light nuclei.
In contrast
to conventional methods of resonance extraction like Argand diagrams and
poles of the $S$-matrix in the complex energy plane, the time delay method
has a physical meaning which was noticed more than 50 years ago by 
Eisenbud \cite{eisen} and Wigner \cite{wigner1}. The fact
that a large positive time delay in a scattering process is associated
with the formation of a resonance is now mentioned and elaborately
discussed in most standard text books \cite{books}. This time delay is 
related to the energy derivative of scattering phase shifts and can be easily 
calculated. However, strangely but not surprisingly, this well-documented
method had rarely been used in literature to extract information on 
resonances from data until recently when the authors of the present work 
put it to a test with hadron-hadron elastic scattering  
\cite{we3KN,we3pi,me,we4}. Being encouraged by the fact that this method was
successful in characterizing known nucleonic resonances \cite{we4}, could find
some evidence for penta-quark baryons \cite{we3KN} and revealed all known 
meson resonances \cite{we3pi}, we carried out a similar study \cite{wejphysg} 
for the $\eta^3$He system which also proved quite fruitful. 
Hence, in the present work we shall search for $\eta$d and $\eta^4$He 
unstable states using the time delay method.  
The transition matrix for $\eta$-nucleus scattering (which is 
required as an input to the calculation of time delay) is 
constructed using few body equations and accounts 
for off-shell re-scattering and 
nuclear binding energy effects. 
This model was successful \cite{umebkj} 
in showing that the enhancement in the cross sections 
(or the strong energy dependence of the scattering amplitude near threshold) 
of the $p d \rightarrow \,^3$He$\,\eta$ reaction is due to the 
$\eta\,^3$He final state interaction. 
Since the calculations are purely theoretical, we have extended
the concept of time delay to negative energies too. In the following
sections, we shall demonstrate the validity and usefulness of this
concept in characterizing bound and virtual states occurring at negative
energies.

There exist scores of papers in literature, with predictions of 
unstable bound (sometimes called quasi-bound) states of 
$\eta$-mesic nuclei as well as $\eta$-nucleus resonances. 
One can find a detailed account of the
existing literature in a recent work \cite{hailiu2}, where a search for 
unstable $\eta$-mesic nuclei ranging from $A=3$ to $208$ was made. Here, we 
shall briefly survey the literature on light nuclei, since we restrict 
ourselves to the study of $\eta$d and $\eta^4$He elastic 
scattering in the present work. One of the early calculations \cite{ueda} 
using the 3-body equation, predicted a resonance with a mass of 
2430 MeV and a width of 10-20 MeV in the $\pi N N$-$\eta N N$ coupled system. 
This state led to a remarkable enhancement of the $\eta d$ elastic 
cross section. In some other Faddeev-type calculations   
\cite{shev,shev2} of $\eta d$ scattering, a resonance 
at low energies was predicted. The exact values of the
predicted mass and the width of course varied with the value of the 
$\eta$-nucleon 
scattering length, $a_{\eta N}$, which in turn depends on the model of
the $\eta$-nucleon interaction. The existence of resonances in the above
works was inferred from Argand diagrams. Some other recent 
(also Faddeev-type) calculations, however, ruled out the possibility of a
resonance in the $\eta d$ system \cite{gpena,delof,fix}. Predictions 
of resonances in the light $\eta$-nucleus systems, namely, $d$, $^3$He and 
$^4$He, from the positions and movements of the poles of the amplitude, can
be found in \cite{rakit96}. Finally to mention some predictions of 
`virtual' states, a virtual (anti-bound) s-wave $\eta ^3$H state, which led
to a large enhancement of the cross section for $\eta$ production from the
three-body nuclei was found in \cite{Fix3N}, whereas a narrow virtual state 
in the $\eta d$ system was found \cite{wycech} to have a rather weak effect on 
the $p n \rightarrow \eta d$ cross sections.

In this work, we shall infer the existence of unstable states in the $\eta$d 
and $\eta ^4$He systems, from positive peaks in time delay and
compare our results with existing predictions in literature using other 
methods. In the next section, we present the basic elements of the ``time
delay" method and the characteristics expected 
from it in the case of resonances, bound, quasi-bound, virtual and 
quasi-virtual states.  
Since one can find thorough discussions of
time delay in literature \cite{alltdpapers,smith},
and also in standard text-books
on quantum mechanics \cite{books,brans}, we do not perform a review of this
method here. However, the efficacy of the inferences from this method 
is established by applying it to the known case of neutron-proton bound
and virtual states and referring to our earlier application to the known 
hadron resonances. Section 3 is devoted to a brief discussion of the 
few body equations involving the finite rank approximation (FRA) which we
use to calculate the $\eta$-nucleus transition matrix. This $t$-matrix is 
subsequently used to evaluate the time delay in $\eta$-light nucleus 
scattering. Since the applicability of the FRA for the $\eta d$ case 
is limited (it was shown in \cite{shev2} that FRA agrees with the 
Faddeev calculations for real part
of the scattering lengths less than $0.5$ fm only), we use a 
parameterization of the $\eta$ d $t$-matrix using 
relativistic Faddeev equations and present the
results of this calculation in Sec. 4. The time delay results for the 
$\eta ^4$He system are given in Sec. 5. 
The FRA calculations are done using two different models of the $\eta N$ 
interaction (constructed within coupled channel formalisms) which fit the
same set of data on pion scattering and pion induced $\eta$ production on a 
nucleon, but give different values of the $\eta N$ scattering length. 
Finally, Sec. 6 
summarizes the findings of the present work.

\section{Time delay plots of bound, virtual and decaying states} 

The collision time or delay time in scattering processes, was quantified 
by Eisenbud and Wigner in terms of the observable phase shift 
which can be extracted from cross section 
data. In \cite{wigner1}, Wigner pictured the resonance formation in 
elastic scattering as the capture and retention of the incident particle 
for some time by the scattering centre, introducing thereby a time delay 
in the emergence of the outgoing particles. 
He further showed that the energy derivative of 
the phase shift, $\delta$, which is related to the time delay, 
$\Delta t(E)$, as
\begin{equation}\label{1}
\Delta t(E) = 2 \hbar {d \delta \over dE}
\end{equation}
and is large and positive close to resonances, can also take negative values 
which are however limited from causality constraints. 
In the presence of inelasticities, a one
to one correspondence between time delay and the lifetime of a resonance 
does not hold and a more useful definition, namely, the time delay matrix 
(later discussed in terms of a lifetime matrix by Smith \cite{smith}) was 
given by Eisenbud. An element of this matrix, $\Delta t_{ij}$, which is the 
time delay in the emergence of a particle in the $j^{th}$ channel after 
being injected in the $i^{th}$ channel is given by, 
\begin{equation}\label{2}
\Delta t_{ij} = \Re e \biggl [ -i \hbar (S_{ij})^{-1} {dS_{ij} \over dE}
\biggr ] \, ,
\end{equation}
where $S_{ij}$ is an element of the corresponding $S$-matrix. 
Writing the $S$-matrix in terms of the $T$-matrix as,
\begin{equation}\label{6}
{\bf S} = 1 + 2\,i\,{\bf T}\, ,
\end{equation}
one can evaluate time delay in terms of the $T$-matrix. 
The time delay in elastic scattering, i.e. $\Delta t_{ii}$, is given in terms 
of the $T$-matrix as,  
\begin{equation}\label{7}
S^*_{ii} \,S_{ii}\, \Delta t_{ii}\, =\, 2 \,\hbar\,
\biggl[ \Re e \biggl({dT_{ii} \over dE}\biggr)\,+ \,2 \,\Re e T_{ii}\,\,
\Im m \biggl ({dT_{ii} \over dE}\biggr) \,-\, 2\, \Im m T_{ii}\,\,
\Re e\biggl( {dT_{ii} \over dE}\biggr)\,
 \biggr],
\end{equation}    
where {\bf T} is the complex $T$-matrix such that, 
\begin{equation} \label{7a}
T_{kj} = \Re e T_{kj} \,+ \,i\,\Im m T_{kj}\,.
\end{equation}
The time delay evaluated from eqs (\ref{1}) and (\ref{7}) is just the same. 

The above relations were put 
to a test in \cite{we3KN,we3pi,me,we4} to characterize the hadron resonances
occurring in meson-nucleon and meson-meson elastic scattering. The energy 
distribution of the time delay evaluated in these works, nicely displayed 
the known $N$ and $\Delta$ baryons, meson resonances like the 
$\rho$, the scalars ($f_0$) and strange $K^*$'s found in $K \pi$ scattering, 
in addition to confirming 
some old claims of exotic states. The peaks in time delay, $\Delta t(E)$, 
agreed well with the known resonance masses. It was also shown that the
time delay peaks and the $T$-matrix poles essentially contain the same 
information. A theoretical discussion on this issue can be found in 
\cite{brans,peres}.

The time delay concept is not only useful to locate resonances, but can
also be used to locate the bound, virtual and unstable bound states 
which have negative binding energies. 
We illustrate this assertion with the well-known case of the $n-p$ system. 
The $S$-matrix for the neutron-proton system, constructed from a square
well potential which produces the correct binding
energy of the deuteron is given as a function of $l$ as,
\begin{equation}
S_l = - {\alpha h_l^{(2)'}(\alpha) j_l(\beta) - \beta h_l^{(2)}(\alpha)
 j_l{'}(\beta) \over 
\alpha h_l^{(1)'}(\alpha) j_l(\beta) - \beta h_l^{(1)}(\alpha) j_l^{'}(\beta)}
\end{equation}
where $j_l$, $h_l^{(1)}$ and $h_l^{(2)}$ are the spherical Bessel and Hankel
functions of the first and second kind respectively.
\begin{equation}
\alpha = k R \,\,\,\,\,{\rm and}\,\,\,\,\, 
\beta = (\alpha^2 - 2 \mu U R^2/\hbar^2)^{1/2}
\end{equation}
where the potential $U$ is given by
\begin{equation}
U = V + i W , \,\,\,\,\, U(r) = U \theta (R-r)
\end{equation}
and $R$ is the width of the potential well.
A similar square well potential was used by Morimatsu and Yazaki \cite{mori}
while locating the ``unstable bound states" of $\Sigma$-hypernuclei as
second quadrant poles and by J. Fraxedas and J. Sesma using the time
delay method \cite{frax}.
\begin{figure}[ht]
\centerline{\vbox{
\psfig{file=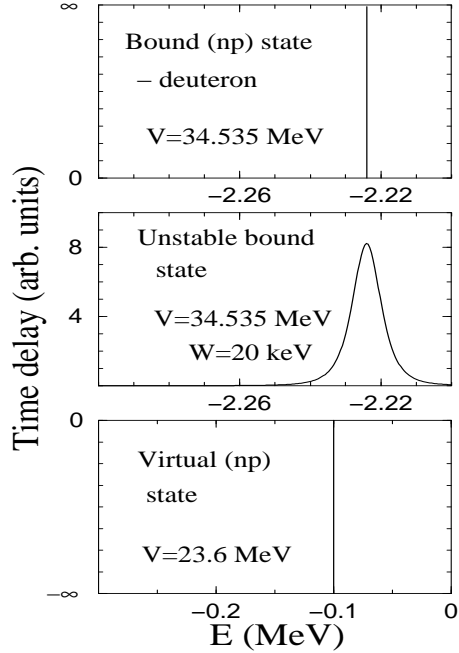,height=9cm,width=6cm}}}
\caption{The theoretical delay time in $n p$
elastic scattering as a function of the energy $E = \sqrt s - m_n - m_p$,
where $\sqrt s$ is the total energy available in the $n p$ centre of
mass system. 
}
\end{figure}
We evaluate the time delay in $n-p$ scattering at negative energies $E$, 
where $E = \sqrt s - m_n - m_p$ with $\sqrt s$ being the energy available 
in the $n p$ centre of mass system.  
Using the above $S$-matrix with $\alpha = i k R$ (hence $E = -k^2/2\mu$), 
$l=0$ and the appropriate parameters for an $n-p$
square well potential, namely, $V = 34.6 MeV$ and $R = 2.07$ fm, the time delay
plot (as in Fig. 1) shows a sharp spike (positive infinite time delay) 
exactly at the
binding energy of the deuteron ($E=-2.224$ MeV). If we add a small
imaginary part to the potential, then of course there is a
Breit-Wigner kind of distribution centered around the binding energy
of the deuteron (a fictitious ``unstable bound state" of the $n-p$
system at $E = -2.224$ MeV).

The correlation between the potential 
parameters and the position of the spike at the correct deuteron
binding energy is very definite. If we take the potential parameters 
which do not give the correct binding energy, then the spike appears
at a wrong place in the time delay plot. 
In Fig. 2, we demonstrate this sensitivity of time delay. 
In the upper half of the figure, we plot time 
delay calculated using a fixed well depth of $V = 34.535$ MeV and 
different choices for the width of the square well. 
It can be seen that the position of the spike is very sensitive to
the value of $R$. The spike at the right binding energy is produced 
only with $R=2.07$ fm. 
In the lower half of the plot, we perform a
similar calculation but this time with the well width fixed and 
the well depth changing. Again, a small change in the well depth 
parameter, changes the energy at which the positive infinite time
delay appears.

\begin{figure}[ht]
\centerline{\vbox{
\psfig{file=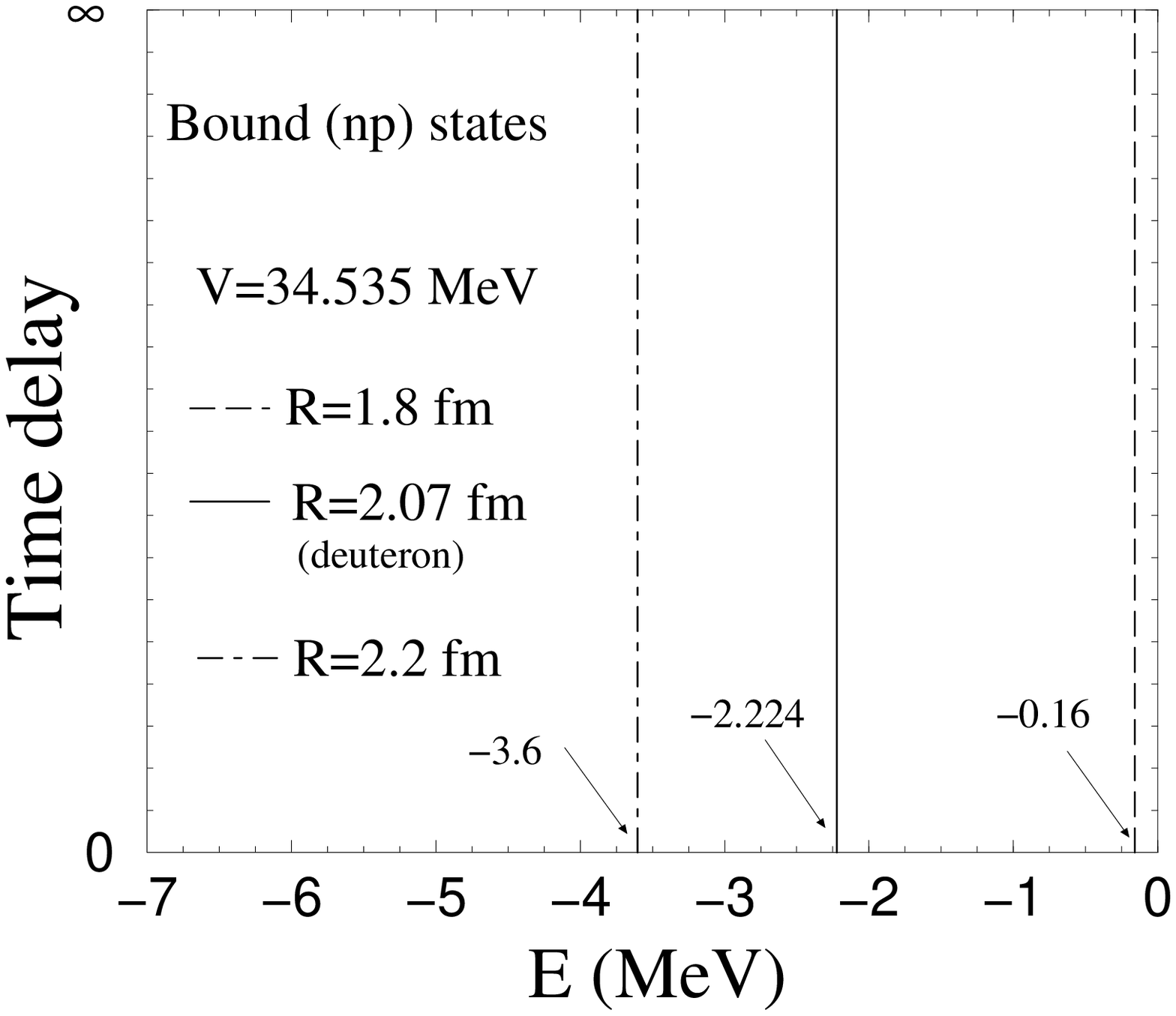,height=5cm,width=6.5cm}
\psfig{file=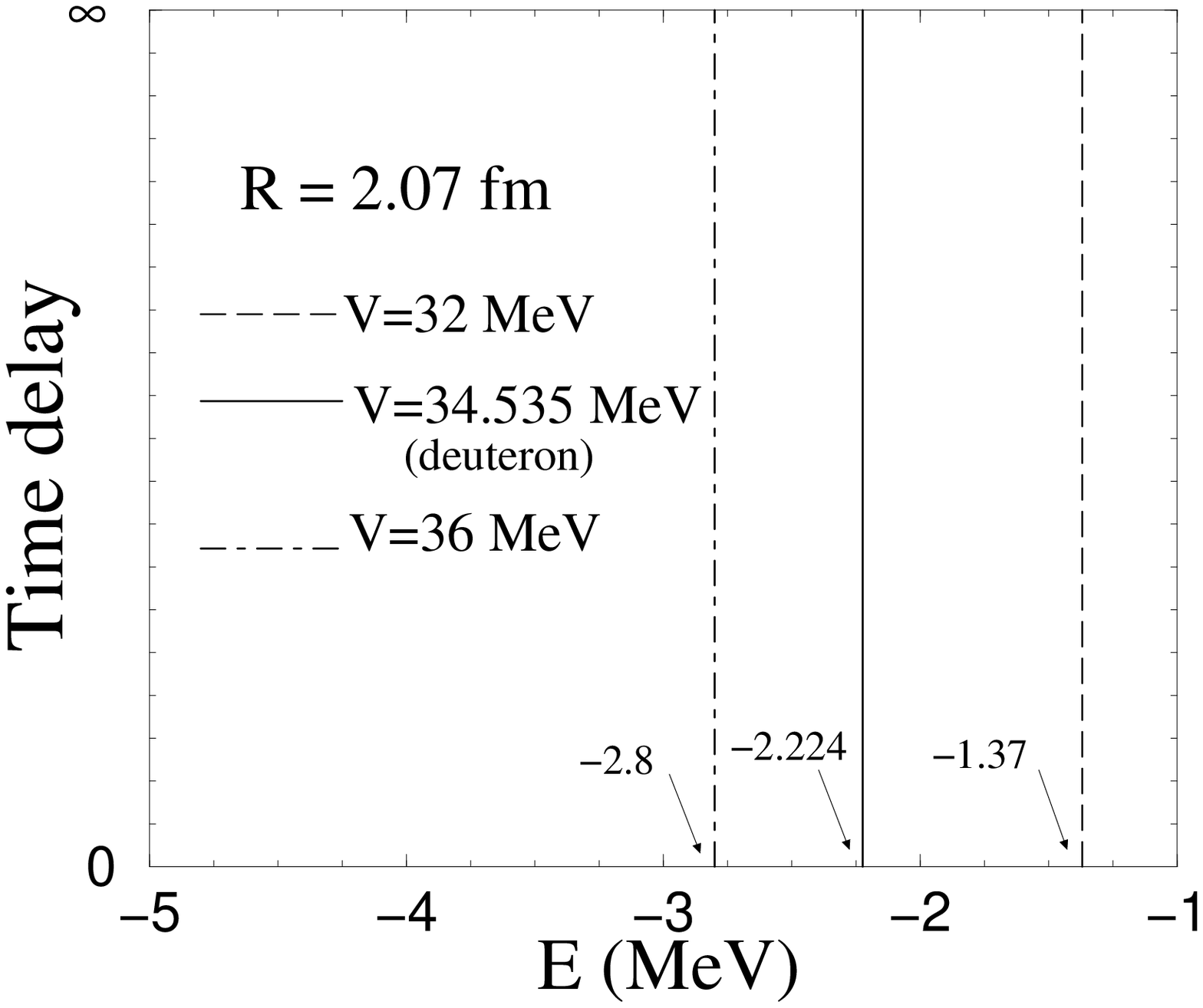,height=5cm,width=6.5cm}}}
\caption{\label{fig:epsart2} Sensitivity of time delay in $n p$ 
scattering to the square well parameters. $V$ is the depth of the
square well potential and $R$ its width. The solid lines
indicate the sharp positive infinite time delay at the correct deuteron binding 
energy.}   
\end{figure}
Beyond quasibound states, an $S$-matrix pole in the third quadrant 
of the complex momentum plane corresponds to a quasivirtual state, 
which translates to a pole on the unphysical sheet
of the complex energy plane of the type $-|E| + i |\Gamma/2|$. 
This, in contrast to a resonance pole of $|E| - i |\Gamma/2|$ (which
leads to an exponentially decaying state with a decay law of 
$e^{-\Gamma t}$), gives rise to an exponential
growth, namely $e^{+\Gamma t}$. One can then see that in contrast to
the time ``delay" that one observes for a resonance, one would observe
a time ``advancement" for a quasivirtual state. In other words, for a 
quasivirtual state one observes a finite `negative' time delay. 
Similarly, a virtual state, in contrast to a bound state would 
show an infinite negative time delay. This is indeed seen in Fig. 1 for
the known $n-p$ virtual state at $100$ keV, where the time delay 
calculated from the square well potential with parameters corresponding
to this virtual state (namely, $V = 23.6$ MeV and $R = 2$ fm) is 
plotted.

\section{$T$-matrix for $\eta$-nucleus elastic scattering}

We evaluate the transition matrix for $\eta$-nucleus ($\eta A$) 
elastic scattering, using few body equations for the $\eta(2N)$ 
and $\eta(4N)$ systems. The calculation is done within a Finite Rank 
Approximation (FRA) approach, which means that in the intermediate state, 
the nucleus in $\eta A$ elastic
scattering remains in its ground state. Since the $\eta$-mesic bound states
and resonances are basically low energy phenomena, it seems justified to use
the FRA for calculations of the present work. 
In \cite{shev2}, the authors mention that though
the use of FRA for $\eta^3$He and $\eta^4$He systems seems justified, it is
questionable for the case of $\eta d$ scattering and investigate
the shortcoming due to the neglect of excitations of the nuclear ground state 
in $\eta$-deuteron calculations. Within their model, they find that the 
FRA results differ from those evaluated using the rigorous 
Alt-Grassberger-Sandhas (AGS) equations in the case of the strong $\eta N$ 
interaction (Re $a_{\eta N} > 0.5$ fm), while for small values of 
Re $a_{\eta N}$, the FRA is reasonably good. 
Therefore, for the $\eta$-d case, we also include calculations using 
the results from the recent relativistic Faddeev equation (RFE) calculations 
for the $\eta N N$ system.

The target Hamiltonian $H_A$, in the FRA is written as \cite{bela},
\begin{equation}\label{hamilt}
H_A \approx \varepsilon |\psi_0> <\psi_0|
\end{equation}
where $\psi_0$ is the nuclear ground state wave function and
$\varepsilon$
the binding energy.
The $\eta A$ $T$-matrix in the FRA is given as \cite{rakit96,bela,rakit1},
\begin{eqnarray}\label{tfsi}
t_{\eta A}(\vec{k^\prime},\, \vec{k}\,; z) &=& <\, \vec {k^\prime}\, ; \,
\psi_0\,|\, t^0(z)
\, | \, \vec{k} \, ; \, \psi_0\,> \, +\, \\ \nonumber
&&\varepsilon\, \int {\vec{dk^{\prime\prime}}
\over (2\pi)^3} {<\,\vec{k^\prime}\, ; \,  \psi_0 \,|\, t^0(z)\, | \,
\vec{k^{\prime\prime}}\,
; \,  \psi_0\,> \over (z - {k^{\prime\prime\,2} \over 2\mu})(z -
\varepsilon
- {k^{\prime\prime\,2} \over 2\mu})}
t_{\eta A}(\vec{k^{\prime\prime}},\, \vec{k}\, ; \, z)
\end{eqnarray}
where $z = E - |\varepsilon| + i0$. $E$ is the energy associated with
$\eta A$ relative motion, $\varepsilon$ is the binding energy of the 
nucleus and $\mu$ is the reduced mass of the
$\eta A$ system. Though the operator $t^0$ describes the
scattering of the $\eta$ meson from nucleons fixed in their space position
within the nucleus, it differs from the usual fixed center $t$-matrices. 
Here, $t^0$ is taken off the energy shell and involves the 
motion of the $\eta$ meson with respect to the center of mass of the
target. The present scheme should not be confused with a conventional
optical potential approach which involves the impulse approximation
and omits the re-scattering of the $\eta$ meson from the nucleons. 
The matrix elements for $t^0$ are given as,
\begin{equation}\label{t0mat}
<\, \vec{k^\prime} \, ; \,  \psi_0\,|\, t^0(z)\, |\, \vec {k} \, ; \,
\psi_0\,> =
\int d\vec{r}\, |\, \psi_0(\vec{r})\, |^2 \, t^0\, (\vec{k^\prime},\,
\vec{k}\,;
\vec{r}\,;z)
\end{equation}
where,
\begin{equation}\label{t0mat2}
t^0\,(\vec{k^\prime},\, \vec{k} \,;\vec{r} \,;z) = \sum_{i=1}^A \,
t_i^0\,
(\vec{k^\prime},\, \vec{k}\,;\vec{r_i}\,;z)
\end{equation}
$t_i^0$ is the t-matrix for the scattering of the $\eta$-meson from the
$i^{th}$ nucleon in the nucleus, with the re-scattering
from the other (A-1) nucleons included. It is given as,
\begin{equation}\label{t0mat3}
t_i^0\,(\vec{k^\prime},\, \vec{k}\,;\vec{r_i}\,;z) =
t_i^{\eta N}(\vec{k^\prime},\, \vec{k}\,;\vec{r_i}\,;z) + \int
{d\vec{k^{\prime\prime}}
\over (2\pi)^3}\,{t_i^{\eta N}(\vec{k^\prime},\,
\vec{k^{\prime\prime}}\,;\vec{r_i}\,;z)
\over z - {k^{\prime\prime\,2} \over 2\mu}} \sum_{j\neq i}
t_j^0(\vec{k^{\prime\prime}},\, \vec{k}\,;\vec{r_j}\,;z)
\end{equation}
The t-matrix for elementary $\eta$-nucleon scattering, $t_i^{\eta N}$, 
is written in terms of the two body $\eta N$ matrix 
$t_{\eta N \rightarrow \eta N}$ as,
\begin{equation}\label{tetan}
t_i^{\eta N}(\vec{k^\prime},\, \vec{k}\,;\vec{r_i}\,;z) =
t_{\eta\,N \rightarrow \eta N}(\vec{k^\prime},\,
\vec{k}\,;z)\, exp [\,i (\, \vec{k} -
\vec{k^\prime}\,)\cdot\,\vec{r_i}\,]
\end{equation}
The $^4$He nuclear wave function, required in the
calculation of the $T$-matrix is taken to be of the Gaussian form. 
The deuteron wave function is written using a parametrization of
the wave function \cite{paris} obtained using the Paris potential. The results
using the Paris potential are also compared with a calculation using a 
Gaussian form of the deuteron wave function.

As mentioned in the introduction, there exists a lot of uncertainty 
in the knowledge of the $\eta$-nucleon interaction and hence, 
we use different prescriptions of
the $\eta$-N t-matrix, t$_{\eta \, N \, \rightarrow \, \eta \, N}$, 
leading to different values of the $\eta N$ scattering length. We give
a brief description of two of these models of 
t$_{\eta \, N \, \rightarrow \, \eta \, N}$ which we use for the FRA 
calculations below. In  
\cite{fix} a coupled channel t-matrix including the $\pi$N and 
$\eta$N channels with the S$_{11}$ - $\eta$N interaction playing a 
dominant role was constructed. The t-matrix thus consisted of the 
meson - N* vertices and the N* propagator as given below:
\begin{equation}\label{tfix}
t_{\eta \, N \, \rightarrow \, \eta \, N} (\, k^\prime, \, k; z) = 
{ { \rm g}_{_{N^*}}\beta^2 \over (k^{\prime\,2} +
\beta^2)}\,\tau_{_{N^*}}(z)\,{ {\rm g}_{_{N^*}}\beta^2 \over (k^2 + \beta^2)}
\end{equation}
with,
\begin{equation}\label{tau}
\tau_{_{N^*}}(z) = ( \, z - M_0- \Sigma_\pi(z) - 
\Sigma_\eta(z) + i\epsilon \, )^{-1}
\end{equation}
where $\Sigma_\alpha(z)$ $(\alpha = \pi, \eta)$ are the self energy
contributions from the $\pi N$ and  $\eta N$ loops. 
We choose the parameter set with ${\rm g}_{_{N^*}} = 2.13, \,\beta = 13 \,{\rm fm}^{-1},\,\,\,{\rm and} 
\,\,\,M_0 = 1656 \,{\rm MeV} $ which leads to $a_{\eta N}$ = (0.88, 0.41) fm.

We also present results using one of the earliest calculations of 
the $\eta$-N t-matrix \cite{bhale} which gives
a much smaller value of the scattering length, 
namely, $a_{\eta N}$ = (0.28, 0.19) fm. 
In this model, 
the $\pi$N, $\eta$N and $\pi \Delta$ ($\pi \pi N$) channels were treated 
in a coupled channel formalism (so that an additional self-energy 
term appears in the propagator in Eq. (\ref{tau})). 
The parameters of this model are, 
$ { \rm g}_{_{N^*}} = 0.616, \,\beta = 2.36 \,{\rm fm}^{-1},\,\,\,{\rm and} 
\,\,\,M_0 = 1608.1 \,{\rm MeV}.$

There also exists a recent model of the $\eta N$ interaction 
\cite{green2}, which
predicts a scattering length of $a_{\eta N} = (0.91, 0.27)$ fm, from a
fit to the $\pi N \to \pi N$, $\pi N \to \eta N$, $\gamma N \to \pi N$
and $\gamma N \to \eta N$ data. However, we have not used it for our present
FRA calculations since the $T$-matrix which fits the data very well is
an on-shell $T$-matrix.
The off-shell separable form given by the authors  \cite{green2}
agrees with their on-shell $T$-matrix (which fits data) but does
not include the intermediate off shell $\pi$ and $\eta$ loops. 
The off shell nature appears only in the vertex form factors.

The $T$-matrix for $\eta A$ elastic scattering, $t_{\eta A}$, is related to
the $S$-matrix as,
\begin{equation}\label{smat1}
S\,=\, 1 \,-\, {\mu \,i\,k \over \pi}\,t_{\eta A}
\end{equation}
where $k$ is the momentum in the $\eta A$ centre of mass system 
and hence, the dimensionless $T$-matrix required in the evaluation of 
time delay as given in eq. (\ref{7}) is evaluated using the relation, 
\begin{equation}\label{ttd}
T\,=\,-\,{\mu \,k \over \,2\,\pi} \,t_{\eta A}
\end{equation}
We shall present the time delay plots for $\eta$d and 
$\eta\,^4$He elastic scattering in the next sections.

\section{The $\eta$ deuteron system}
We make an analytic continuation of the $T$ matrix for $\eta$-nucleus
elastic scattering on to the complex energy plane. Evaluating the
matrix elements of the $\eta$-nucleus $T$-matrix at 
negative energies (corresponding
to purely imaginary momentum), i.e. 
$t_{\eta A}(\vec{ik},\, i\vec{k}\,; z)$, we evaluate the
time delay in $\eta$-nucleus elastic scattering and 
search for the ``unstable bound states".  
The resonances at positive energies are of course determined from
the positive time delay peaks at positive energies and real momenta.
\begin{figure}[ht]
\centerline{\vbox{
\psfig{file=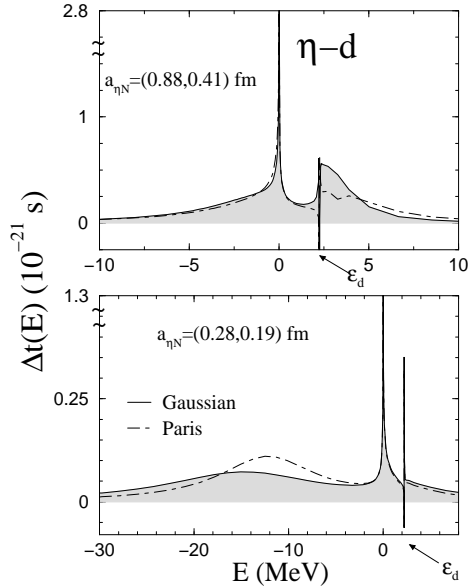,height=8cm,width=6cm}}}
\caption{The delay time in $\eta$-deuteron 
elastic scattering as a function of the energy $E = \sqrt s - m_{\eta} - m_d$,
where $\sqrt s$ is the total energy available in the $\eta d$ centre of
mass system. The shaded curves are calculations using the Gaussian form 
of the deuteron wave function and dashed lines are evaluated using the
Paris deuteron wave function. The vertical axis scale is broken in order
to display the structure in the time delay plot clearly.}
\end{figure}

In Fig. 3, we plot the time delay in $\eta$ d $\rightarrow \eta$ d 
elastic scattering with two different inputs for the elementary 
$\eta N$ interaction. 
The $\eta N$ scattering lengths of 
$a_{\eta N} = (0.88, 0.41)$ fm and $a_{\eta N} = (0.28, 0.19)$ fm
correspond to $\eta$d scattering lengths of 
$a_{\eta d} = (1.52, 2.57)$ fm and $a_{\eta d} = (0.67, 0.42)$ fm 
respectively. 
In both cases, we see a large positive time delay
located near threshold. The solid lines (with shaded regions)
are the calculations using a Gaussian form and the dashed lines
with a Paris potential parametrization of the deuteron wave function.
The vertical axis scale in Fig. 3 is broken in order
to display the structure in the time delay plot clearly.
We do not find a big difference in the results with the change of
the wave function. 
In the case of the weaker $\eta N$ interaction, i.e. 
$a_{\eta N} = (0.28,0.19)$ fm, there appears a very broad bump
around $-15$ MeV which could be due to an unstable bound
state.  
On the positive energy side, there is a sharp negative time delay when 
the kinetic energy
of the $\eta$-d system equals the binding energy of the deuteron.
This behaviour is expected because of the connection of the 
energy derivative of the phase shift and hence the time delay 
(\ref{1}) to the density of states as given by the 
Beth-Uhlenbeck formula \cite{uhlen}. 
It was shown in \cite{me} that a maximum negative time delay occurs at 
the opening of an inelastic threshold. 
In the present case, the negative dip around $2.22$ MeV, corresponds to 
the break up threshold of the deuteron. 
We also see a resonance just near this inelastic threshold. 
However, the above result near the inelastic threshold should be taken
with some caution since we have used the FRA which might not be a very
good approximation at energies where new thresholds open up.

\begin{figure}
\centerline{\vbox{
\psfig{file=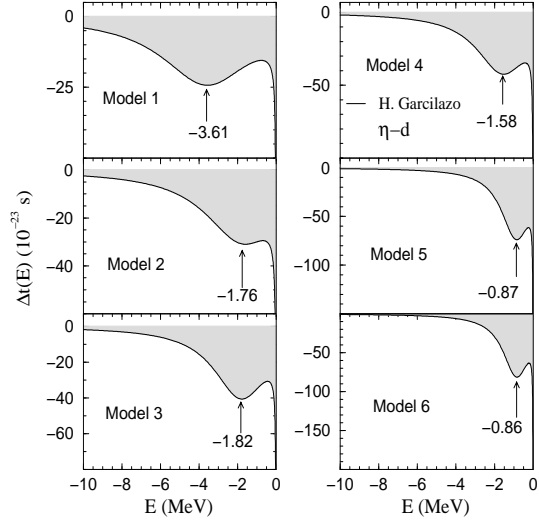,height=7cm,width=7cm}}}
\caption{Quasivirtual states of the $\eta d$ system 
evaluated from a Faddeev calculation for $\eta d$ scattering. The various
model numbers correspond to the different strengths of the $\eta N$ 
interaction as explained in \cite{garcil}.}
\end{figure}
\begin{figure}
\centerline{\vbox{
\psfig{file=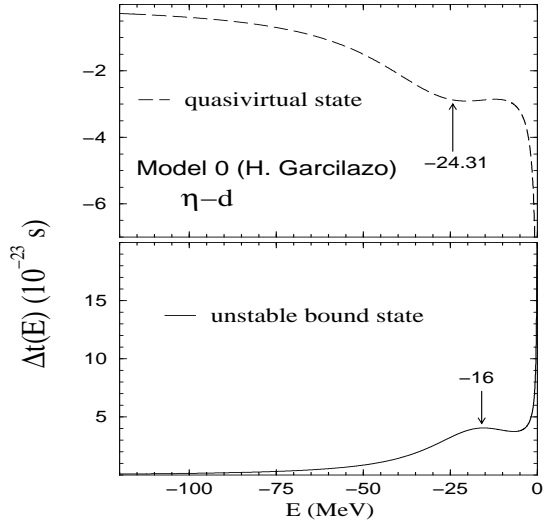,height=7cm,width=7cm}}}
\caption{Quasivirtual and quasibound states 
of the $\eta d$ system 
evaluated from a Faddeev calculation for $\eta d$ scattering using
model 0 as mentioned in \cite{garcil}.}
\end{figure}
In order to check the validity of the above FRA approach 
for the $\eta d$ case (where it is known to have limitations \cite{shev2}), 
we evaluate time delay 
using a model \cite{garcil} which obtains the $\eta d$ elastic 
scattering amplitude using a relativistic version of the Faddeev 
equations described in \cite{garcil2}. The $\eta d$ amplitude in 
\cite{garcil} is parametrized using the effective range formula:
\begin{equation}
f^{-1}_{\eta d} = {1 \over A_{\eta d}} \, + \, {1 \over 2} 
\, R_{\eta d} \, k^2 \, + \, S_{\eta d}\, k^4\, -i\,k
\end{equation}
where $k$ is the momentum in the $\eta d$ centre of mass system 
and the parameters $A_{\eta d}$, $S_{\eta d}$ and $R_{\eta d}$ are
as given in Table II of \cite{garcil}. 
In Figs 4 and 5, we show the results obtained using the above 
amplitude. The time delay in Fig. 4 has been evaluated using 
$k \to -ik$ and hence the negative dips in this figure correspond
to the quasivirtual states which appear as poles 
in the third quadrant of the complex k-plane. The locations of
these dips (using different models as given in \cite{garcil}) 
are exactly at the energy pole values given in Table III 
of \cite{garcil} as expected. As explained already in Sec. II, 
such a negative time delay (or time advancement) is expected for
quasivirtual states which give an exponential rise rather than an 
exponential decay law.

\begin{figure}
\centerline{\vbox{
\psfig{file=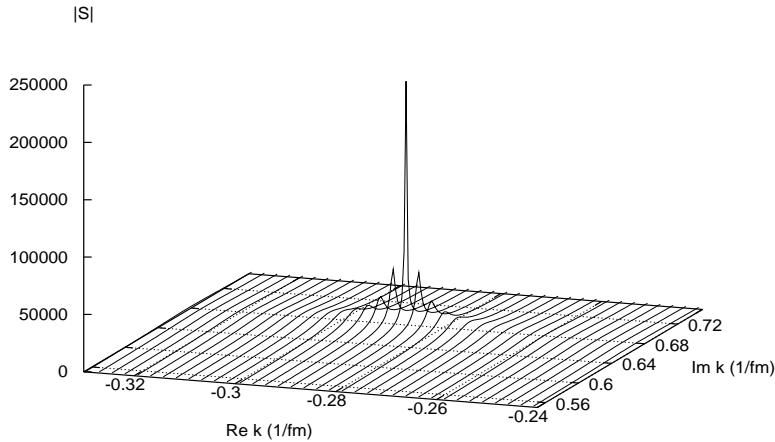,height=8cm,width=12cm}}}
\caption{Magnitude of the complex amplitude 
$S= 1+ 2ikf$ in the complex momentum plane 
evaluated from a Faddeev calculation for $\eta d$ scattering using
model 0 as mentioned in \cite{garcil}. The sharp pole corresponds to
the quasibound state at $-16$ MeV as seen in the time delay plot in Fig. 5.}
\end{figure}
In the upper half of Fig. 5, we plot the time delay evaluated using 
$k \to -ik$ and the lower half shows the time delay plot
corresponding to $k \to +ik$ in the amplitude with Model 0. 
Thus the negative
dip in the upper half is the quasivirtual state as also given in
Table II of \cite{garcil} and the lower half shows the positive  
time delay corresponding to a quasibound or what we address as
an ``unstable bound" state in the present work. In order to 
demonstrate once again the one-to-one correspondence between the
time delay peak and $S$-matrix poles, in Fig. 6 we plot the 
magnitude of $S$ as a function of the real and imaginary parts 
of momentum $k$. The pole occurs at ($-0.283 + i0.674$) fm 
in the complex momentum plane. This corresponds to a peak value
of about $-17$ MeV and a width of $35$ MeV which is in good agreement
with the time delay plot in Fig. 5.  
Such a pole has however not been mentioned by the author in 
\cite{garcil} (probably due to the fact that the author in \cite{garcil}
has been mostly concerned about the effect of the near threshold 
quasivirtual states on the $n p \rightarrow d \eta$ reaction).
We do not find any such positive peaks in any of the models shown 
in Fig. 4. It is interesting to note that as expected from \cite{shev2},
the FRA has some agreement with the Faddeev calculation for small
$\eta N$ scattering lengths, $a_{\eta N}$, 
and the discrepancy increases for large $a_{\eta N}$.

Finally, we wish to caution the reader regarding
the interpretation of time delay peaks in the case of $s$-wave scattering.
To see this, substituting the phase shift expression, 
$S = exp(2i\delta)$ and comparing it with (\ref{smat1}), one
can write,
\begin{equation}
\delta = {1 \over 2i} \, {\rm ln}(1 - {i \mu k \over \pi} t_{\eta A}) \, = \,
{1 \over 2i} \, {\rm ln}(1 + 2 i k f) 
\end{equation}
where $f$ is the scattering amplitude. For small $k$, $\delta \simeq k f$ and
the behaviour of $d\delta/dE$ (the real part of
which is essentially the time delay) is
determined by the simple pole at $k=0$ (or $E_{\eta A} = E_{threshold}$) and
the energy dependence of the scattering amplitude $f$. In the absence of
a resonance, as $k \to 0$, $\delta = k a$, where $a = a_R + i a_I$
is the complex scattering length. For positive energies,
$\Re e \delta = k a_R$, whereas for energies below zero, $ k \to ik$ and
$\Re e \delta = - k a_I$. In such a situation, $\Re e (d\delta/dE)$ exhibits
a sharp peak at $E_{\eta A} = E_{threshold}$, the sign of which is determined 
by the sign of the scattering length.
On the other hand, if the scattering amplitude has a resonant behaviour
near threshold, one would see a superposition of the two behaviours.
Consequently, a state reasonably close
to threshold gets distorted in shape and one very close
manifests simply by broadening the threshold singularity. A state
far from threshold, however, remains completely unaffected.

\section{The $\eta \,^4$He  system}

\begin{figure}[ht]
\centerline{\vbox{
\psfig{file=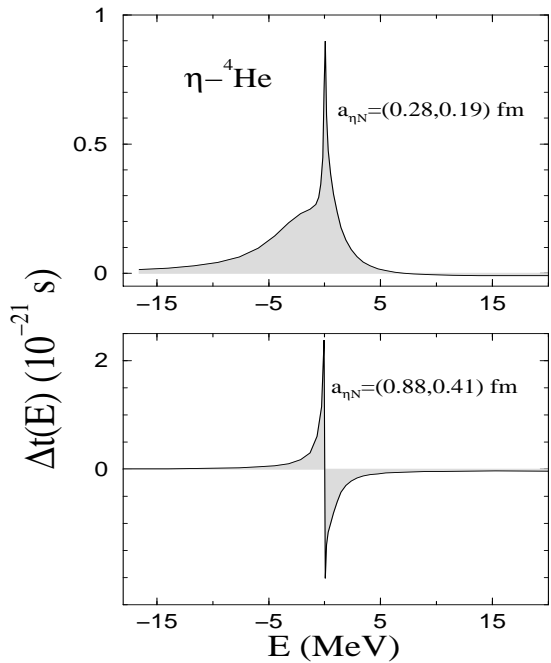,height=9cm,width=7cm}}}
\caption{The delay time in $\eta \,^4$He 
elastic scattering as a function of the energy $E = \sqrt s - m_{\eta} - 
m_{\eta ^4He}$,
where $\sqrt s$ is the total energy available in the $\eta \,^4$He centre of
mass system.}  
\end{figure}
The time delay plot in the upper half of Fig. 7, 
shows once
again the near threshold peak and one more broad one centered
around $-2$ MeV. However, the plot with time delay evaluated using 
the model which gives $a_{\eta N} = (0.88, 0.41)$ fm, shows a
large negative time delay near threshold. The negative time delay
can sometimes also arise due to a repulsive interaction \cite{smith}.
Intuitively, an attractive interaction is something that causes
resonance formation and ``delays" the scattering process. A repulsive
interaction on the other hand would speed up the process and the
time taken for the process in the absence of interaction would be larger
than that with interaction.

Before ending this section, we note that the $\eta^4$He  
scattering lengths obtained within the FRA and 
using two different models of the $\eta N$ interaction, namely,
$a_{\eta N} = (0.88, 0.41)$ fm and $a_{\eta N} = (0.28, 0.19)$ fm
are $a_{\eta \,^4He} = (-3.94, 5.575)$ fm and 
$a_{\eta \,^4He} = (1.678, 1.524)$ fm respectively.

\section{Summary}
In conclusion, we summarize the present work as: \\
(i) we have made a search for the 
unstable states of $\eta$-mesic deuteron and $^4$He 
extending the approach of time delay 
which was used recently for the first time in the eta-mesic 
case \cite{wejphysg}.\\
(ii) The established time delay method for searching resonances has 
been extended to negative energies, to search for bound, virtual and unstable 
bound states. The validity of this method has been established by 
applying it first to the known case of the $n p$ system and then
to the case of the $\eta$-d system within a parameterized 
Faddeev calculation.\\    
(iii) The calculations were performed
with different values of the $\eta N$ scattering length considered
as acceptable in literature.
Within the Faddeev equation parameterization, we find one unstable bound state 
far from threshold ($\sim -16 $ MeV) for an $\eta N$ 
scattering length of $(0.42,0.34)$ fm.
Within the FRA calculation, we find such an $\eta d$ state 
around $-12$ MeV for $a_{\eta N} = (0.28, 0.19)$ fm.
These results seem to indicate that though the FRA in general is not
recommendable for $\eta d$ elastic scattering, the results are close to
those from the Faddeev calculations for low values of the $\eta N$
scattering length.
\\
(iv) In the $\eta^4$He case, within the FRA calculations, we find
an unstable bound state close to threshold 
for a small scattering length of $(0.28,0.19)$ fm.

\end{document}